   \newcommand{\be}[0]{\begin{equation}}
   \newcommand{\ee}[0]{\end{equation}}
   \newcommand{\ba}[0]{\begin{eqnarray}}
   \newcommand{\ea}[0]{\end{eqnarray}}    
\documentstyle[12pt]{article}
\begin{document}
\Large
\hfill\vbox{\hbox{DTP/00/12} \hbox{February 2000}}
            
\nopagebreak

\vspace{0.75cm}
\begin{center}
\LARGE
{\bf Complete Renormalization Group Improvement-  Avoiding Factorization 
and Renormalization Scale Dependence in QCD Predictions}
\vspace{0.6cm}
\Large

C.J. Maxwell$^{1)}$ and A. Mirjalili$^{2)}$

\vspace{0.4cm}
\large
\begin{em}
Centre for Particle Theory, University of Durham\\
South Road, Durham, DH1 3LE, England
\end{em}

\vspace{1.7cm}

\end{center}
\normalsize
\vspace{0.45cm}

\centerline{\bf Abstract}
\vspace{0.3cm}
%%%%%%%%%%%%ABSTRACT%%%%%%%%%%%%%%%%%%
For moments of leptoproduction structure functions we show that all 
dependence on the renormalization and factorization scales disappears 
provided that all the ultraviolet logarithms involving the physical 
energy scale $Q$ are completely resummed.   
The approach is closely related to Grunberg's method of
Effective Charges. 
A direct and simple method for extracting ${\Lambda}_{\overline{MS}}$
from experimental data is advocated.

\vfill\footnoterule\noindent
$^1$) C.J.Maxwell@durham.ac.uk\\
$^2$) Abolfazl.Mirjalili@durham.ac.uk

\newpage

\section*{1 Introduction}
%%%%%%%%%%%%A:INTRODUCTION%%%%%%%%%%%%%%%%%%

The problem of renormalization scheme dependence in QCD perturbation theory
remains on obstacle to making precise tests of the theory. In a recent paper \cite{r1}
one of us pointed out that the renormalization scale dependence of dimensionless 
physical QCD observables, depending on a single energy scale $Q$, can be avoided 
provided that all ultraviolet logarithms which build the physical energy 
dependence on $Q$ are resummed. This was termed complete Renormalization Group
(RG)-improvement in Ref.\cite{r1}. It was stressed that standard
RG-improvement, as customarily applied with a $Q$-dependent scale ${\mu}=xQ$
, omits an infinite subset of these logarithms. One should rather keep ${\mu}$
independent of $Q$, and then carefully resum to all-orders the RG-predictable ultraviolet
logarithms. In this way all ${\mu}$-dependence cancels between the 
renormalized coupling and the logarithms of ${\mu}$ contained in the coefficients,
and the correct physical $Q$-dependence is built. At next-to-leading order (NLO)
the result is identical to the Effective Charge approach of Grunberg
\cite{r1a,r1b}.
We wish to extend this argument to processes involving factorization of 
operator matrix elements and coefficient functions, where a factorization scale $M$
arises in addition to the renormalization scale $\mu$.
We shall use the prototypical factorization problem of moments of 
leptoproduction structure functions as a specific example. We shall identify the logarithms of $\mu$
,$M$ , and $Q$  which occur, and will show explicitly that on resumming all the 
ultraviolet logarithms the $\mu$ and
$M$ dependence disappears.
We shall organize the paper so that we review the treatment of Ref.[1]
whilst showing how it generalizes for the moment problem. We begin in Section 2
by giving some basic definitions for the moments of structure functions.
Section 3 considers the dependence of the perturbative coefficients on the
parameters which label the renormalization procedure in both cases. Section 4
deals with the complete RG-improvement of the structure function moments
and identifies and resums the physical ultraviolet logarithms. Finally, in
Section 5 we discuss a more straightforward way of motivating this
approach, and consider how to directly extract ${\Lambda}_{\overline{MS}}$ from
data. We also give our Conclusions.

\section*{2 Structure Function Moments} 
%%%%%%%%%%%%B:Considering the Moment%%%%%%%%%%%%%%%%%%
In the prototypical factorization problem of deep inelastic leptoproduction
the $n^{\rm{th}}$ moment of a non-singlet structure function $F(x)$, 
\be
{{\cal{M}}_n(Q)}=\int_{0}^{1}x^{n-2}F(x)\,dx\;,
\ee
can be factorized in the form
\be
{{\cal{M}}_n(Q)}=<{{\cal{O}}_n(M)}>{{\cal{C}}_n(Q,{{a}(\mu)},\mu,M)}\;.
\ee
Here $M$ is an arbitrary factorization scale and
$a(\mu)$ is the RG-improved coupling ${\alpha_s}({\mu})/{\pi}$ defined at a 
renormalization scale $\mu$.
The operator matrix element $<{{\cal{O}}_n(M)}>$ has an $M$-dependence 
given by its anomalous dimension, 
\be
\frac{M}{<\cal{O}>}\frac{{\partial}{<\cal{O}>}}{{\partial}{M}}   
={{\cal{\gamma}}_{\cal{O}}
(a)}=-da-{d_1}{a^2}-{d_2}{a^3}-{d_3}{a^4}+\ldots\;
\ee
For simplicity we shall from now on suppress the $n$-dependence of terms in  equations,
as we have done in Eq.(3). For a given moment $d$ is independent of
the factorization convention, whereas the higher $d_i$ ,($i{\ge}1$)
depend on it.
In Eq.(3) the coupling $a$ is governed by the $\beta$-function equation
\be
{M}\frac{{\partial}a}{{\partial}{M}}=   
{{\cal{\beta}}(a)}=-b{a^2}(1+ca+{c_2}{a^2}+{c_3}{a^3}+{\ldots})\;.
\ee
Here $b=(33-2{N_f})/6$ , and $c=(153-19{N_f})/12b$ , are the first two coefficients
of the beta-function for SU$(3)$ QCD with $N_f$ active flavours of quark. They are
universal, whereas the subsequent coefficients ${c_2},{c_3},\ldots$ are scheme-dependent.
Equation (3) can be integrated to \cite{r2,r3}
\be
<{{\cal{O}}(M)}>=A\exp[{\int_{0}^{a}}{\frac{{{\cal{\gamma}}(x)}}
{{{\cal{\beta}}(x)}}}\,dx-{\int_{0}^{\infty}}{\frac{{{\cal{\gamma}}^{(1)}(x)}}
{{{\cal{\beta}}^{(2)}(x)}}}\,dx]\;,
\ee
where ${\gamma}^{(1)}$ and ${\beta}^{(2)}$ denote these functions truncated at one
and two terms, respectively.
The factor $A$ is scheme-independent
\cite{r3} and can be fitted to experimental data.
The second integral in Eq.(5)  is an infinite constant of integration.
In Eq.(2)  ${{\cal{C}}(Q,{{a}(\mu)},\mu,M)}$ is the coefficent function and has the 
 perturbation series
\be
{{\cal{C}}(Q,{\tilde{a}},\mu,M)}=1+{r_1}{\tilde{a}}+{r_2}{{\tilde{a}}^2}+
{r_3}{{\tilde{a}}^3}+{\ldots}\;.
\ee
We shall use ${\tilde{a}}$ to stand for $a(\mu)$ and ${a}$ for $a(M)$.
After combining the integrals in Eq.(5) one obtains
\be
{\cal{M}}=A{\left(\frac{ca}{1+ca}\right)}^{d/b}\;\exp({\cal{I}}(a))\;
(1+{r_1}{\tilde{a}}+{r_2}{\tilde{a}}^{2}+{r_3}{\tilde{a}}^{3}+\ldots)\;,
\ee
where  ${\cal{I}}(a)$ is the finite integral
\be
{\cal{I}}(a)={\int
_{0}^{a}}{dx}\;\frac{{d_1}+({d_1}c+{d_2}-d{c_2})x+({d_3}+c{d_2}
-{c_3}d){x^2}+\ldots}{b(1+cx)(1+cx+{c_2}{x}^{2}+{c_3}{x}^{3}+\ldots)}\;,
\ee
which can be readily evaluated numerically. The coupling $a({\tau})$ itself,
where ${\tau}{\equiv}b\ln({\mu}/{\tilde{\Lambda}})$ , is obtained as the solution
of the transcendental equation \cite{r4} 
\be
\frac{1}{a}+c{\ln}\frac{ca}{1+ca}={\tau}-{\int_{0}^{a}}{dx}\left[-\frac{1}{B(x)}
+\frac{1}{{x^2}(1+cx)}\right]\;,
\ee
where $B(x){\equiv}{x^2}(1+cx+{c_2}{x^2}+{c_3}{x^3}+\ldots)$.

\section*{3 RS and FS dependence of the coefficients}
%%%%%%%%%%%%%%%%%%%%%%C:RS and FS dependence of the coefficients%%%%%%
We first wish to parametrize the dependence of the ${r_n}$ in the coefficient 
function on the renormalization scheme (RS) and factorization scheme (FS).
 
Recall first \cite{r4} that for the single scale case of a dimensionless observable 
${\cal{R}}(Q)$ with perturbation series 
\be
{{\cal{R}}(Q)}=a+{r_1}{a^2}+{r_2}{a^3}+\ldots+{r_n}{a^{n+1}}
+\ldots\;,
\ee
the RS can be labelled by the non-universal coefficients of the beta-function 
${c_2}$,${c_3}$,...,and by ${\tau}$, 
which can be traded as a parameter for ${r_1}$ since \cite{r1a,r1b,r4,r4a}
\be
{\tau}-{r_1}={{\rho}_{0}}(Q){\equiv}b\ln(Q/{{\Lambda}_{\cal{R}}})
\;,
\ee
is an RS-invariant.Using the self-consistency of perturbation theory- that is 
that the difference between a ${\rm{N}}^n$LO calculation (i.e up to and 
including ${r_n}{a}^{n+1}$) performed with two different RS's is 
O(${a}^{n+2}$), one can derive expressions for the partial derivatives of the 
perturbative coefficients with respect to the scheme parameters. For ${r_2}$ for 
instance one has \cite{r4}
\ba
{\frac{\partial{r_2}}{\partial{r_1}}}=2{r_1}+c,\;
{\frac{\partial{r_2}}{\partial{c_2}}}=-1,\;
{\frac{\partial{r_2}}{\partial{c_3}}}=0,...  .
\ea
on integration one finds
\ba
{r_2}({r_1},{c_2})&=&{r_1}^{2}+c{r_1}+{X_2}-{c_2}
\nonumber\\
{r_3}({r_1},{c_2},{c_3})&=&{r_1}^{3}+{5\over2}c{r_1}^{2}
+(3{X_2}-2{c_2}){r_1}+{X_3}-{1\over2}{c_3}
\nonumber\\
\vdots & &\vdots\;.
\ea
In general the structure is
\be
{r_n}({r_1},{c_2},{\ldots},{c_n})={{\hat{r}}_{n}}({r_1},{c_2},
{\ldots},{c_{n-1}})+{X_n}-{c_n}/(n-1)\;,
\ee
where ${\hat{r}}_{n}$ is RG-predictable from a complete 
${\rm{N}}^{n-1}$LO calculation (i.e. ${r_2},{r_3},{\ldots},{r_n}$, and $ {c_2},{c_3},
{\ldots},{c_n}$  have been computed in some RS),
 and the  $X_n$ are $Q$-independent and RS-invariant
constants of integration which are unknown unless a complete  ${\rm{N}}^{n}$LO calculation
has been performed.\\

As we  shall see the generalization to the moment problem is a dependence
${r_n}({\mu},M,
{c_2},{\dots},{c_n},{d_1},{d_2},{\ldots},{d_n})$
where the ${c_i}$ label the RS and the ${d_i}$ the FS. As before $M$,${\mu}$ can
be traded, in this case for ${r_1}(M)$ and ${\tilde{r}_{1}}{\equiv}{r_1}(M={\mu})$.
There will be analogous factorization and renormalization scheme (FRS) invariants
,${X_n}$, which represent the RG-unpredictable parts of ${r_n}$. Expressions
for the dependence of the coefficients on FRS parameters have been derived before in
Refs.\cite{r2,r3,r4b} , but there were some errors in Ref.\cite{r2}, in particular the
dependence of ${r_2}$ on $c_2$ was not recognized \cite{r3}. 
Partially differentiating Eq.(7) with respect to ${\mu},{M},{c_2},{c_3},{d_1},{d_2}
,{d_3}$,  and demanding for consistency that it be O($a^4$) , so that the coefficients
of $a
,{a^2}$ and $a^3$ vanish, one obtains analogous to Eqs.(12), 
\ba
{\mu}\frac{{\partial}{r_1}}{{\partial}{\mu}}&=&0,\;\;\;{\mu}\frac{{\partial}{r_2}}
{{\partial}{\mu}}={r_1}b ,\;\;\; {\mu}\frac{{\partial}{r_3}}{{\partial}{\mu}}
=2{r_2}b+{r_1}bc\;,
\nonumber \\
M\frac{{\partial}{r_1}}{{\partial}M}&=&d,\;\;\;M\frac{{\partial}{r_2}}
{{\partial}M}={d_1}+d{r_1}-dL\;,
\nonumber \\
M\frac{{\partial}{r_3}}{{\partial}M}&=&{d_2}+{d_1}{r_1}+d{r_2}
-d{r_1}L-2{d_1}L-d{L^2},
\nonumber \\
\frac{{\partial}{r_1}}{{\partial}{d_1}}&=&-\frac{1}{b}\;,\frac{{\partial}{r_2}}
{\partial{d_1}}=\frac{c}{2b}-\frac{L}{b}-\frac{r_1}{b},
\nonumber \\
\frac{{\partial}{r_3}}{{\partial}{d_1}}&=&\frac{c{r_1}}{2b}-\frac{c^2}{3b}+
\frac{(c-{r_1})}{b}L-\frac{{r_2}}{b}+\frac{{c_2}}{3b}-\frac{L^2}{b},
\nonumber \\
\frac{\partial{r_1}}{\partial{d_2}}&=&0,\;\;\;\frac{\partial{r_2}}{\partial{d_2}}
=-\frac{1}{2b},\;\;\;\frac{\partial{r_3}}{\partial{d_2}}=\frac{c}{3b}-\frac{L}{b}-
\frac{r_1}{2b}\;,
\nonumber \\
\frac{\partial{r_1}}{\partial{d_3}}&=&0,\;\;\;\frac{\partial{r_2}}{\partial{d_3}}=0
,\;\;\;\frac{\partial{r_3}}{\partial{d_3}}=-\frac{1}{3b}\;,
\nonumber \\
\frac{\partial{r_1}}{\partial{c_2}}&=&0,\;\;\;\frac{\partial{r_2}}{\partial{c_2}}
=\frac{3d}{2b},\;\;\;\frac{\partial{r_3}}{\partial{c_2}}=\frac{4{d_1}}{3b}+
3\frac{dL}{b}+3\frac{d{r_1}}{2b}-{r_1}-5\frac{cd}{3b}\;,
\nonumber \\
\frac{\partial{r_1}}{\partial{c_3}}&=&0,\;\;\;\frac{\partial{r_2}}{\partial{c_3}}
=0,\;\;\;\frac{\partial{r_3}}{\partial{c_3}}=\frac{5d}{6b}\;.
\ea
Here we have defined for convenience $L{\equiv}b{\ln}(M/{\mu})$. Consistently
integrating the partial derivatives of ${r_1}$ yields
\be
{r_1}=
\frac{d}{b}{{\tau}_M}-\frac{d_1}{b}-{X_1}(Q)\;,
\ee
where ${\tau}_{M}{\equiv}b{\ln}(M/{\tilde{\Lambda}})$ and ${X_1}(Q)$ is an
FRS-invariant, analogous to ${\rho}_{0}(Q)$ for the single scale problem
defined in Eq.(11). Exactly analogous to ${\Lambda}_{\cal{R}}$ , for the
moment problem one can define an FRS-invariant ${\Lambda}_{\cal{M}}$
so that
\be
\frac{d}{b}{\tau}_{M}-\frac{{d_1}}{b}-{r_1}={X_1}(Q){\equiv}d{\ln}\left(\frac{Q}
{{\Lambda}_{\cal{M}}}\right)\;.
\ee
Consistently integrating the remaining partial derivatives and using Eq.(16)
to recast the $M$ and ${\mu}$ dependence in terms of ${r_1}$ and ${\tilde{r}}_{1}$,
one obtains the explicit dependence of ${r_2}$ and ${r_3}$ on the FRS
parameters ${r_1},{\tilde{r}}_{1},{d_1},{d_2},{d_3},{c_2},{c_3}$ ,
\setlength\arraycolsep{2pt}
\ba
{r_2}&=&(\frac{1}{2}-\frac{b}{2d}){r_1}^{2}+\frac{b}{d}{r_1}{\tilde{r}}_{1}
+\frac{c{d_1}}{2b}-\frac{d_2}{2b}-\frac{d{c_2}}{2b}+{X_2}
\nonumber \\
{r_3}&=&(\frac{b^2}{d^2}-\frac{3b}{2d}+\frac{1}{2})\frac{{r_1}^3}{3}+
(-\frac{b^2}{d^2}+\frac{b}{d}){r_1}^{2}{\tilde{r}}_{1}+
(\frac{bc}{d}+\frac{2b{d_1}}{d^2}){r_1}{\tilde{r}}_{1}
\nonumber\\
&&+(-\frac{bc}{2d}-\frac{b{d_1}}{d^2}+\frac{d_1}{d}){r_1}^{2}
+(-\frac{d{c_2}}{2b}+\frac{c{d_1}}{2b}+{X_2}+\frac{{d_1}^2}{2db}+\frac{d_2}{d}
-\frac{d_2}{2b}-{c_2}){r_1}
\nonumber\\
&&+(\frac{{d_1}^2}{d^2}-\frac{d_2}{d}+\frac{c{d_1}}{d}+\frac{2b{X_2}}{d})
{\tilde{r}}_{1}+\frac{b^2}{d^2}{r_1}{{\tilde{r}}_{1}}^{2}+(-\frac{{d_1}{c^2}}{3b}
+\frac{2{d_1}{X_2}}{d}
\nonumber\\
&&+\frac{{d_1}^{3}}{3b{d^2}}+\frac{dc{c_2}}{3b}+\frac{c{d_1}^{2}}{2db}+
\frac{d_3}{3d}-\frac{d{c_3}}{6b}-\frac{2{d_1}{c_2}}{3b}+\frac{{d_2}c}{3b}+{X_3})
\nonumber\\
\vdots & & \vdots \;,
\ea
analogous to Eqs.(13) in the single scale case. Notice that we could equally
use $r_1$ and $L$ as parameters instead of $r_1$ and ${\tilde{r}}_{1}$, since
$L=({b/d})({r_1}-{\tilde{r}}_{1})$. As in the single scale case there are
constants of integration $X_n$ representing the RG-unpredictable part
of $r_n$. They are $Q$-independent and FRS-invariant.\\

In the single scale case parametrizing the RS-dependence using
${r_1},{c_2},{c_3},{\ldots}$ means that given a complete ${\rm{N}}^{n}$LO calculation
${X_2},{X_3},{\ldots},{X_n}$ will be known. Using Eqs.(13) to sum
to all-orders the RG-predictable terms, i.e. those {\it not} involving
${X}_{n+1},{X}_{n+2},{\ldots}$, with coupling $a({r_1},{c_2},{c_3},{\ldots})$
is equivalent to ${\rm{N}}^{n}$LO perturbation theory in the scheme
with ${r_1}={c_2}={c_3}={\ldots}=0$, and yields the sum
\be
{\cal{R}}^{(n)}={a_0}+{X_2}{a_0}^{2}+{X_3}{a_0}^{3}+{\ldots}+{X_n}{a_0}^{n}\;,
\ee
where ${a_0}{\equiv}a(0,0,0,{\ldots})$ is the coupling in this scheme. From
Eqs.(9) and (11)  it satisfies
\be
\frac{1}{a_0}+c{\ln}\left(\frac{c{a_0}}{1+c{a_0}}\right)=b{\ln}\left(
\frac{Q}{{\Lambda}_{\cal{R}}}\right)\;.
\ee
In fact the solution of this transcendental equation can
be written in closed form in terms of the Lambert $W$-function \cite{r4c,r5},
defined implicitly by $W(z)\exp(W(z))=z$,
\ba
{a_0}&=&-\frac{1}{c[1+W(z(Q))]}
\nonumber \\
z(Q)&{\equiv}&-\frac{1}{e}{\left(\frac{Q}{{\Lambda}_{\cal{R}}}\right)}^{-b/c}\;.
\ea 
A similar expansion to Eq.(19), but motivated differently, has been suggested
in Ref.\cite{r5a}.\\

In the moment problem by an exactly similar argument, with the chosen
parametrization of FRS, given a complete ${\rm{N}}^{n}$LO calculation 
(i.e. a calculation of ${r_1},{r_2},{\ldots},{r_n}$ and the ${d_1},{d_2},{\ldots},{d_n}$
and ${c_2},{c_3},{\ldots},{c_n}$ in some FRS) the invariants ${X_2},{X_3},{\ldots},{X_n}$
will be known. Using Eqs.(18) to sum to all-orders the RG-predictable
terms not involving ${X}_{n+1},{X}_{n+2},{\ldots}$ , will be equivalent to
working with an FRS in which all the FRS parameters are set to zero.
${\tilde{r}}_{1}=0$ means that ${\mu}={M}$. Setting ${r_1}=0$, ${d_1}=0$
in Eq.(17) yields ${\tau}_{M}=b{\ln}(Q/{\Lambda}_{\cal{M}})$, so that
$a={\tilde{a}}={a_0}$, given by Eq.(21) with ${\Lambda}_{\cal{R}}$ 
replaced by ${\Lambda}_{\cal{M}}$. Further, with ${c_i}={d_i}=0$
the integral ${\cal{I}}(a)$ in Eq.(8) vanishes, so that finally the
sum of all RG-predictable terms for the moment problem at ${\rm{N}}^{n}$LO
will be
\be
{\cal{M}}=A{\left(\frac{c{a_0}}{1+c{a_0}}\right)}^{d/b}(1+{X_2}{a_0}^{2}
+{X_3}{a_0}^{3}+{\ldots}+{X_n}{a_0}^{n})\;,
\ee
with an extremely similar structure to the single scale case in Eq.(19).
Substituting for ${a_0}$ in terms of the Lambert $W$-function using Eq.(21)
we then obtain
\ba
{\cal{M}}&=&A{[-W(z(Q))]}^{b/d}(1+{X_2}{a_0}^{2}+{\ldots})
\nonumber \\
z(Q)&{\equiv}&-\frac{1}{e}{\left(\frac{Q}{{\Lambda}_{\cal{M}}}\right)}^{-b/c}\;.
\ea
So that moments of structure functions have a $Q$-dependence
naturally involving a power of the Lambert $W$-function.\\

As stressed in Ref.\cite{r1} the result of resumming all RG-predictable terms
depends on the chosen parametrization of RS. By simply translating the
parameters to a new set ${\breve{r}}_{1}={r_1}-{\overline{r}}_{1},
{\breve{c}}_{2}={c_2}-{\overline{c}}_{2},{\ldots}$ etc., where the barred
quantities are constants, one finds corresponding new constants of
integration ${\breve{X}}_{n}$. The result of resumming all RG-predictable
terms with this new parametrization then corresponds to standard fixed-order
perturbation theory in the RS with ${r_1}={\overline{r}}_{1}$,${c_2}={\overline{c}}_{2}
,{\ldots}$,
or equivalently with ${\breve{r}}_{1}={\breve{c}}_{2}={\breve{c}}_{3}={\ldots}=0$.
The key point is that ${r_1}$ has a special status since it
contains the ultraviolet (UV) logarithms which build the physical
$Q$-dependence of ${\cal{R}}(Q)$. Standard RG-improvement corresponds to
shifting the parameter ${r_1}$, in which case the resulting constants
of integration ${\breve{X}}_{n}$ contain physical UV logarithms which are
not all resummed. Thus $r_1$ should be used as the parameter. An exactly
similar statement holds for $r_1$ and ${\tilde{r}}_{1}$ in the moment problem.
We shall identify the UV logarithms and show how their complete resummation
builds the correct physical $Q$-dependence in the next section. \\

We shall refer to the expansions in Eqs.(19) and (22) as Complete RG-improved
(CORGI) results. Whilst the parameters implicitly containing the UV logarithms
do have a special status, the remaining dimensionless parameters $c_i$ and $d_i$ can be
reparametrized as one pleases. As an example, in the Effective Charge
approach of Grunberg \cite{r1a,r1b} one chooses ${\overline{c}}_{2},{\overline{c}}_{3},{\ldots},
{\overline{c}}_{n}$ so that ${\breve{X}}_{2},{\breve{X}}_{3},{\ldots},{\breve{X}}_{n}$
are all zero at ${\rm{N}}^{n}$LO, corresponding to ${r_1}={r_2}={\ldots}={r_n}=0$, and
this is {\it a priori} equally reasonable. In the moment problem one can
correspondingly choose the ${\overline{c}}_{i}$ and ${\overline{d}}_{i}$ so that
at ${\rm{N}}^{n}$LO the ${\breve{X}}_i$ all vanish and ${r_1}={r_2}={\ldots}={r_n}=0$.
If one further demands that the integral ${\cal{I}}(a)$ in Eq.(8) vanishes order-by-order
in $a$ a unique FRS is selected in which moments have the form
\be
{\cal{M}}=A{\left(\frac{c{\hat{\cal{R}}}}{1+c{\hat{\cal{R}}}}\right)}^{d/b}\;.
\ee
Where ${\hat{\cal{R}}}$ is an effective charge which has a perturbation series
of the form,
\be
{\hat{\cal{R}}}=a+{\hat{r}}_{1}{a}^{2}+{\hat{r}}_{2}+{\ldots}+{\hat{r}}_{n}{a}^{n+1}+{\ldots}\;.
\ee
This is similar to Grunberg's proposal \cite{r1b} to associate an effective charge with
${\cal{M}}$ so that ${\cal{M}}=A{(c{\hat{\cal{R}}})}^{d/b}$. The ${\hat{r}}_{i}$ are
built from the ${c_i},{d_i},M $ and ${\mu}$,and are RS-dependent, but FS-independent.
Effectively ${\hat{\cal{R}}}$ can be RG-improved as in the single scale case.
We have for instance
\be
{\hat{r}}_{1}=b{\ln}({\mu}/{\tilde{\Lambda}})-b{\ln}({M}/{\tilde{\Lambda}})-\frac{b}{d}{r_1}
+{d_1}/{d}= {\tau}-{X_1}(Q)\;,
\ee
where we have used Eq.(17) . Comparing with Eq.(11) we see that treating ${\hat{\cal{R}}}$
as a single scale problem we have ${\rho}_{0}(Q)={X_1}(Q)$. This further implies that
${\Lambda}_{\hat{\cal{R}}}={\Lambda}_{\cal{M}}$ and so the corresponding CORGI
couplings are identical. The CORGI expansion for ${\hat{\cal{R}}}$ will be of the
form
\be
{\hat{\cal{R}}}={a_0}+{\hat{X}}_{2}{a_0}^{2}+{\hat{X}}_{3}{a_0}^{3}+{\ldots}\;.
\ee
Inserting this result in Eq.(24) and re-expanding in $a_0$ will reproduce the
CORGI expansion in Eq.(22).

\section*{4 Complete RG-improvement}
%%%%%%%%%%%%D:Resummation of series%%%%%%%%%%%%%
In the single scale case using Eq.(11) one can write
\be
{r_1}=b\left({\ln}\frac{\mu}{\tilde{\Lambda}}-{\ln}\frac{Q}{{\Lambda}_{\cal{R}}}\right)\;.
\ee
The first ${\mu}$-dependent logarithm depends on the RS, whereas the second
$Q$-dependent UV logarithm will generate the physical $Q$-dependence and is
RS-invariant. If one makes the simplification that $c=0$ and sets
${c_2}={c_3}=...=0$, then the coupling is given by
\be
a(\mu)=1/b{\ln}(\frac{\mu}{{\tilde{\Lambda}}})\;.
\ee
The sum to all-orders of the RG-predictable terms from Eqs.(13), given a
NLO calculation of ${r_1}$, simplifies to a geometric progression,
\be
{\cal{R}}=a+{r_1}{a}+{{r_1}^2}{a^3}+\ldots+{{r_1}^{n}}{a^{n+1}}+\ldots\;.
\ee
The idea of complete RG-improvement is that dimensionful renormalization
scales, in this case $\mu$, should be held strictly independent of the
physical energy scale $Q$ on which ${\cal{R}}(Q)$ depends. In this way the
$Q$-dependence is built entirely by the ``physical'' UV logarithms
$b{\ln}(Q/{\Lambda}_{\cal{R}})$ contained in ${r_1}$, and the convention-dependent
logarithms of ${\mu}$ cancel between $a(\mu)$ and ${r_1}({\mu})$ ~, when the
all-orders sum in Eq.(30) is evaluated. The conventional fixed-order
NLO truncation ${\cal{R}}=a(\mu)+{r_1}({\mu}){a(\mu)}^2$, only makes sense
if ${\mu}=xQ$, but then the resulting $Q$-dependence involves the arbitrary
parameter $x$. In contrast using Eqs.(28),(29) and summing the geometric
progression in Eq.(30) gives,
\be
{\cal{R}}(Q){\approx}a(\mu)/\left[1-\left(b{\ln}\frac{\mu}{{\tilde{\Lambda}}}-
b{\ln}\frac{Q}{{\Lambda}_{\cal{R}}}\right)a(\mu)\right]=1/b{\ln}(Q/
{\Lambda}_{\cal{R}})\;,
\ee
correctly reproducing the large-$Q$ behaviour of ${\cal{R}}(Q)$,
\be
{\cal{R}}(Q){\approx}1/b{\ln}(Q/{\Lambda}_{\cal{R}})+O{(1/b{\ln}(Q/{\Lambda}_{\cal{R}})}
^{3}\;.
\ee

In the moment problem the analogous UV logarithm is $b{\ln}(Q/{{\Lambda}_{\cal{M}}})$
introduced in Eq.(17), and analogous to Eq.(28) we will have
\be
{r_1}=d\left({\ln}\frac{M}{{\tilde{\Lambda}}}-{\ln}\frac{Q}{{\Lambda}_{\cal{M}}}
\right)-\frac{{d_1}}{b}\;.
\ee
Given a NLO calculation of $r_1$ we wish to see how the physical $Q$-dependence
of ${\cal{M}}(Q)$ arises on resumming to all-orders the UV logarithms
contained in the RG-predictable terms from Eqs.(18). If we make similar
approximations, so that $c=0$ and the $d_i$ and $c_i$ are set to zero,
then 
\be
{\cal{M}}=A{(ca(M))}^{d/b}(1+{r_1}a({\mu})+{r_2}{a(\mu)}^{2}+\ldots)\;.
\ee
We retain the overall factor of $c^{d/b}$. The task is then to show
that on resumming the RG-predictable terms in the coefficient function
to all-orders the ${\ln}(M/{\tilde{\Lambda}})$ and ${\ln}({\mu}/{\tilde{\Lambda}})$
contained in $r_1$ and ${\tilde{r}}_{1}$ cancel with those in the couplings
$a(M)$ and $a({\mu})$ to yield the physical Q-dependence
\be
{\cal{M}}(Q){\approx}A{c}^{d/b}{(1/b{\ln}(Q/{\Lambda}_{\cal{M}}))}^{d/b}
(1+O{(1/{\ln}(Q/{\Lambda}_{\cal{M}}))}^{2})\;.
\ee
Again, the complete RG-improvement 
summing over all UV logarithms is forced on one if $\mu$ and $M$ are held
independent of $Q$.\\

The algebraic structure of the resummation of RG-predictable terms for
the moment problem is considerably more complicated than the geometric
progression of Eq.(30) encountered in the single scale case. With
the simplifications $c=0,{c_i}=0,{d_i}=0$ the first two RG-predictable
coefficients from Eqs(18) are
\ba
\setlength\arraycolsep{2pt}
{r_2}&=&({\frac{1}{2}}-{\frac{b}{2d}}){r_1}^{2}+{\frac{b}{d}}{r_1}{\tilde{r_1}}
\\
{r_3}&=&({\frac{b^{2}}{d^{2}}}-{\frac{3b}{2d}}+{\frac{1}{2}})
{\frac{{r_1}^{3}}{3}}+({\frac{-b^{2}}{d^{2}}}+{\frac{b}{d}})
{r_1}^{2}{\tilde{r_1}}+{\frac{b^{2}}{d^{2}}}{r_1}{\tilde{r_1}^{2}}
\ea
Suitably generalizing the partial derivatives in Eqs.(15) one can arrive 
at a general form for the RG-predictable terms. It is useful to
arrange them in columns,
\ba
\left(\begin{array}{cccc}
{r_1}\to({\frac{b}{d}}{\tilde{r_1}})^{0}{r_1}{\tilde{a}} & 0 & 0 & \dots\\
{r_2}\to({\frac{b}{d}}{\tilde{r_1}})^{1}{r_1}{\tilde{a}}^{2} & 
(1-{\frac{b}{d}}){\frac{{r_1}^{2}}{2}}{\tilde{a}}^{2} & 0 & \dots\\
{r_3}\to({\frac{b}{d}}{\tilde{r_1}})^{2}{r_1}{\tilde{a}}^{3} & 
2({\frac{b}{d}}{\tilde{r_1}})
(1-{\frac{b}{d}}){\frac{{r_1}^{2}}{2}}{\tilde{a}}^{3} & 
(1-\frac{b}{d})(\frac{1}{2}-\frac{b}{d}){\frac{{r_1}^{3}}{3}}
{\tilde{a}}^{3} & \dots\\
{r_4}\to({\frac{b}{d}}{\tilde{r_1}})^{3}{r_1}{\tilde{a}}^{4} & 3({\frac{b}{d}}
{\tilde{r_1}})^{2}
(1-{\frac{b}{d}}){\frac{{r_1}^{2}}{2}}{\tilde{a}}^{4} & 
3({\frac{b}{d}}{\tilde{r_1}})(1-\frac{b}{d})(\frac{1}{2}-\frac{b}{d})
{\frac{{r_1}^{3}}{3}}{\tilde{a}}^{4} & \dots\\
{r_5}\to({\frac{b}{d}}{\tilde{r_1}})^{4}{r_1}{\tilde{a}}^{5} &
4({\frac{b}{d}}{\tilde{r_1}})^{3}
(1-{\frac{b}{d}}){\frac{{r_1}^{2}}{2}}{\tilde{a}}^{5} &
6({\frac{b}{d}}{\tilde{r_1}})^{2}(1-\frac{b}{d})(\frac{1}{2}-\frac{b}{d})
{\frac{{r_1}^{3}}{3}}{\tilde{a}}^{5} & \dots\\                
\vdots & \vdots & \vdots & \ddots
\end{array}\right)
\ea
The idea will be to resum each column separately. Denoting the sum of the
${\rm m}^{th}$ column by $S_m$, one finds
\ba 
\setlength\arraycolsep{2pt}
&{S_1}&={r_1}{\tilde{a}}+({\frac{b}{d}}{\tilde{r_1}}){r_1}{\tilde{a}}^{2}+
({\frac{b}{d}}{\tilde{r_1}})^{2}{r_1}{\tilde{a}}^{3}+
({\frac{b}{d}}{\tilde{r_1}})^{3}{r_1}{\tilde{a}}^{4}+
({\frac{b}{d}}{\tilde{r_1}})^{4}{r_1}{\tilde{a}}^{5}+{\ldots}\;
\nonumber\\
&&={r_1}{\tilde{a}}[1+({\frac{b}{d}}{\tilde{r_1}}{\tilde{a}})+
({\frac{b}{d}}{\tilde{r_1}}{\tilde{a}})^{2}+
({\frac{b}{d}}{\tilde{r_1}}{\tilde{a}})^{3}+
({\frac{b}{d}}{\tilde{r_1}}{\tilde{a}})^{4}
+{\dots}\;]
\nonumber\\
&&={r_1}{\tilde{a}}(1-{\frac{b}{d}}{\tilde{r_1}}{\tilde{a}})^{-1}
\ea
Careful examination of the pattern of terms in Eq.(38) leads to the
general result for $S_m$ for $m>1$,
\be
{S_m}=(-1)^{2m-1}(\frac{b}{d}-1)(\frac{b}{d}-{\frac{1}{2}})
(\frac{b}{d}-{\frac{1}{3}})+{\dots}\;+(\frac{b}{d}-{\frac{1}{m-1}})
{\frac{{S_1}^{m}}{m}}
\ee
Finally the resummed RG-predictable terms in the coefficient
function will follow from ${\cal{C}}=1+{S_1}+{S_2}+{S_3}+{\ldots}+{S_n}+{\ldots}$.
Introducing for convenience $x{\equiv}{S_1}={r_1}{\tilde{a}}{(1-\frac{b}{d}{\tilde{r}}_{1}
{\tilde{a}})}^{-1}$
, we find
\ba
\setlength\arraycolsep{2pt}
&{\cal{C}}&=1+x-(\frac{b}{d}-1){\frac{x^{2}}{2}}+
(\frac{b}{d}-1)(\frac{b}{d}-{\frac{1}{2}}){\frac{x^{3}}{3}}
-(\frac{b}{d}-1)(\frac{b}{d}-{\frac{1}{2}})(\frac{b}{d}-{\frac{1}{3}}){\frac{x^{4}}{4}}+{\dots}\;
\nonumber\\
&&=1+\frac{d}{b}(\frac{bx}{d})+\frac{1}{2!}\frac{d}{b}(\frac{d}{b}-1){(\frac{bx}{d})}^{2}+
+\frac{1}{3!}\frac{d}{b}(\frac{d}{b}-1)(\frac{d}{b}-2){(\frac{bx}{d})}^{3}+{\ldots}
\nonumber\\
&&={(1+\frac{b}{d}x)}^{d/b}\;.
\ea
Substituting for $x$ yields
\be
{\cal{C}}=\{1+{\frac{b}{d}}[{r_1}{\tilde{a}}
(1-{\frac{b}{d}}{\tilde{r_1}}{\tilde{a}})^{-1}]\}^{{\frac{d}{b}}}
=[{\frac{1-{\frac{b}{d}}{\tilde{r_1}}{\tilde{a}}+
{\frac{b}{d}}{r_1}{\tilde{a}}}{1-{\frac{b}{d}}{\tilde{r_1}}{\tilde{a}}}}]
^{{\frac{d}{b}}}
\ee
We can write the numerator in Eq.(42) as
\be
(1-{\frac{b}{d}}{\tilde{r_1}}{\tilde{a}}+
{\frac{b}{d}}{r_1}{\tilde{a}})=[1+{\tilde{a}}b
({\frac{{r_1}-{\tilde{r_1}}}{d}})]=(1+{\tilde{a}}L)
\ee
Where $L=b{\ln}(M/{\mu})=b({r_1}-{\tilde{r}}_{1})/d$. Since we are setting
$c={c_2}={c_3}={\ldots}=0$ one has ${(1+{\tilde{a}}L)}^{-1}=a/{\tilde{a}}$,
substituting this into Eq.(42) gives
\be
{\cal{C}}=[(1-{\frac{b}{d}{\tilde{r_1}}{\tilde{a}}}){\frac{a}{\tilde{a}}}]^{\frac{-d}{b}}
=[{\frac{(1-{\frac{b}{d}{\tilde{r_1}}{\tilde{a}}})}{\tilde{a}}}a]^{\frac{-d}{b}}
\ee
Since ${\tilde{a}}=a({\mu})=1/{\tau}$ we can rearrange Eq.(16) to obtain
\be
{\tilde{r_1}}={\frac{d}{b}}{\frac{1}{\tilde{a}}}-d{\ln}\frac{Q}{{\Lambda}_{\cal{M}}}\;,
\ee
and substituting this result into Eq.(43) we find
\be
{\cal{C}}={\left(\frac{1}{b{\ln}(Q/{\Lambda_{\cal{M}}})}\right)}^{d/b}{a}^{-d/b}\;.
\ee
 Combining this with the anomalous dimension part ${(ca)}^{d/b}$ we reproduce
 the physical $Q$-dependence of ${\cal{M}}(Q)$ in Eq.(35).
\section*{5 Discussion and Conclusions}
%%%%%%%%%%%%%%%%%%Conclusion%%%%%%%%%%%%%%%%%%
An alternative and more straightforward way of understanding
the CORGI proposal is as follows. Given a dimensionless observable
${\cal{R}}(Q)$, dependent on the single dimensionful scale $Q$, we
clearly must have, on grounds of generalized dimensional analysis \cite{r8}
\be
{\cal{R}}(Q)={\Phi}\left(\frac{\Lambda}{Q}\right)\;,
\ee
where $\Lambda$ is a dimensionful scale, connected with the
universal dimensional transmutation parameter of the theory,
whose definition will depend on the way in which ultraviolet
divergences are removed, ${\Lambda}_{\overline{MS}}$ for instance.
We can try to invert Eq.(47) to obtain
\be
\frac{\Lambda}{Q}={\Phi}^{-1}({\cal{R}}(Q))\;,
\ee
where ${\Phi}^{-1}$ is the inverse function. This is indeed
the basic motivation for Grunberg's Effective Charge approach \cite{r1a,r1b}.
We are assuming massless quarks here. The extension if one includes
masses has been discussed in \cite{r1b,r9}.
The structure of ${\Phi}^{-1}$ is \cite{r10,r11}
\be
{\cal{F}}({\cal{R}}(Q)){\cal{G}}({\cal{R}}(Q))={\Lambda}_{\cal{R}}/Q\;,
\ee
where 
\be
{\cal{F}}({\cal{R}}(Q)){\equiv}{e}^{-1/b{\cal{R}}}{(1+1/b{\cal{R}})}^{c/b}
\ee
is a universal function of ${\cal{R}}$. ${\Lambda}_{\cal{R}}$ is
connected with the universal parameter ${\Lambda}_{\overline{MS}}$
by the relation
\be
{\Lambda}_{\cal{R}}={e}^{r/b}{\tilde{{\Lambda}}_{\overline{MS}}}\;,
\ee
which follows from Eq.(11), with $r{\equiv}{r_1}^{\overline{MS}}({\mu}=Q)$
the NLO $\overline{MS}$ coefficient. Note that $r$ is $Q$-independent.
The tilde over $\Lambda$ reflects the convention assumed in integrating
the beta-function equation to obtain Eq.(9) \cite{r4}, and ${\tilde{{\Lambda}}}_
{\overline{MS}}={(2c/b)}^{-c/b}{\Lambda}_{\overline{MS}}$ in terms of
the standard convention . The function ${\cal{G}}({\cal{R}}(Q))$
has the expansion
\be
{\cal{G}}({\cal{R}}(Q))=1-\frac{X_2}{b}{\cal{R}}(Q)+O({\cal{R}}^{2})+{\ldots}\;.
\ee
Here $X_2$ is the NNLO RS-invariant constant of integration which arises
in Eqs.(13). Assembling all this we finally obtain the desired inverse
relation between ${\cal{R}}$ and ${\Lambda}$, the universal dimensional
transmutation parameter of the theory
\be
Q{\cal{F}}({\cal{R}}(Q)){\cal{G}}({\cal{R}}(Q)){e}^{-r/b}{(2c/b)}^{c/b}={\Lambda}_
{\overline{MS}}\;.
\ee
Notice that all dependence on the subtraction scheme chosen resides
in the single factor ${e}^{-r/b}$, the remainder of the expression
being independent of this choice. This corresponds to the observation
of Celmaster and Gonsalves \cite{r12}, that ${\Lambda}$'s with different
subtraction conventions can be exactly related given a one-loop (NLO)
calculation. If only a NLO calculation has been performed ${\cal{G}}=1$
since $X_2$ will be unknown, so that the best one can do in reconstructing
${\Lambda}_{\overline{MS}}$ is
\be
Q{\cal{F}}({\cal{R}}(Q)){e}^{-r/b}{(2c/b)}^{c/b}={\Lambda}_{\overline{MS}}\;.
\ee
This is precisely the result obtained on inverting the NLO CORGI result
${\cal{R}}={a_0}$ given by Eq.(21).\\

The essential point is that the dimensional transmutation scale
$\Lambda$ is the fundamental object. In contrast the convention-dependent
dimensionful scales ${\mu}$ and ${M}$ are ultimately irrelevant quantities
which cancel out of physical predictions if one takes care to resum
{\it all} of the ultraviolet logarithms that build the physical $Q$-dependence
in association with $\Lambda$. Our purpose has been to indicate that
the unphysical ${\mu}$ and $M$ dependence of conventional fixed-order
perturbation theory reflects its failure to resum all of these RG-predictable
terms. We have analyzed how Eq.(54) is built by explicitly resumming
the convention-dependent logarithms together with the ultraviolet logarithms.
Having done this, however, one can simply use Eq.(53) to test perturbative QCD.
Given at least a NLO calculation for an observable ${\cal{R}}(Q)$ one simply
substitutes the data values into Eq.(53), where $\cal{G}(\cal{R}(Q))$ can include
NNLO and higher corrections if known, and obtains ${\Lambda}_{\overline{MS}}$.
To the extent that remaining higher-order perturbative and possible power corrections
are small, one should find consistent values of ${\Lambda}_{\overline{MS}}$  for
different observables. There is no need to mention $\mu$ or $M$ in this
analysis, let alone to vary them over an {\it ad hoc} range of values.
For the moment problem the result coresponding to Eq.(53) is
\be
Q\overline{\cal{F}}\left(\frac{{\cal{M}}}{A}\right)\overline{\cal{G}}\left(
\frac{\cal{M}}{A}\right){e}^{-{\hat{r}}/b}{(2c/b)}^{c/b}={\Lambda}_{\overline{MS}}\;,
\ee
where $\hat{r}{\equiv}{{\hat{r}}_{1}}^{\overline{MS}}({\mu}=Q)$ is defined in Eq.(26). The
modified functions $\overline{\cal{F}}$ and $\overline{\cal{G}}$ are most
easily obtained by noting that $\hat{\cal{R}}$ in Eq.(24) is directly related
to ${\cal{M}}/A$ and also satisfies Eq.(53). One finds
\ba
\overline{\cal{F}}(x)&=&{\exp}[bc(1-{x}^{-b/d}]
{(1+bc({x}^{-b/d}-1))}^{c/b}
\nonumber \\
\overline{\cal{G}}(x)&=&\left(1-\frac{X_2}{d}\frac{{x}^{b/d}}{c(1-{x}^{b/d})}+{\ldots}\right)\;.
\ea
Where $X_2$ is the NNLO FRS-invariant which arises in Eqs.(18).
The scheme-independent parameter $A$ reflects a physical property of
the operator ${\cal{O}}_{n}$ in Eq.(2). $A_n$ and ${\Lambda}_{\overline{MS}}$
should be fitted simultaneously to the data for ${\cal{M}}_{n}(Q)$ using 
Eq.(55).\\

We hope to report direct fits of data to ${\Lambda}_{\overline{MS}}$ as 
outlined above, for both ${e}^{+}{e}^{-}$ jet observables \cite{r13} and
structure functions and their moments \cite{r14},  in future publications.
\section*{Aknowledgements}
%%%%%%%%%%%%%%%%%%Aknowledgements%%%%%%%%%%%%%%%%%%
A.M. acknowledges the financial support of the Iranian government and also
thanks S.J. Burby for useful discussions. 

\end{document}